\documentclass{aa}
\usepackage[varg]{txfonts}
\usepackage{amsfonts}
\usepackage{amsmath}
\usepackage{nicefrac}
\usepackage{float}
\usepackage{color}
\usepackage{geometry}
\usepackage{graphicx}
\usepackage{subfigure}
\usepackage{textcomp}
\usepackage{pdfpages}
\usepackage{url}
\usepackage{natbib}
\usepackage{hyperref}
\usepackage{prettyref}
\newrefformat{fig}{\text{Fig.~\ref{#1}}}
\newrefformat{eq}{\text{Eq.~\ref{#1}}}
\newrefformat{tab}{\text{Table~\ref{#1}}}
\newrefformat{sec}{\text{Sect.~\ref{#1}}}

\title{DR Tau: Temporal variability of the brightness distribution in the potential planet-forming region\thanks{Based on observations collected at the European Organisation for Astronomical Research 
in the Southern Hemisphere, Chile, under the programs 074.C-0342(A) and 092.C-0726(A,B).}}

\titlerunning{DR Tau: Temporal variability}

\author{R. Brunngräber\inst{1} \and S. Wolf\inst{1} \and Th. Ratzka\inst{2} \and F. Ober\inst{1}}
\institute{Institute of Theoretical Physics and Astrophysics, Christian-Albrechts-Universität zu Kiel, Leibnizstr. 15, 24118 Kiel, Germany\\\email{rbrunngraeber@astrophysik.uni-kiel.de} \and Institute for Physics/IGAM, NAWI Graz, Karl-Franzens-Universität, Universitätsplatz 5/II, 8010, Graz, Austria}

\date{Received  / Accepted}

\abstract{}{We investigate the variability of the brightness distribution and the changing density structure of the protoplanetary disk around DR Tau, a classical T Tauri star. DR Tau is known for its peculiar variations from the ultraviolet (UV) to the mid-infrared (MIR). Our goal is to constrain the temporal variation of the disk structure based on photometric and MIR interferometric data.}{We observed DR Tau with the MID-infrared Interferometric instrument (MIDI) at the Very Large Telescope Interferometer (VLTI) at three epochs separated by about nine years, two months, respectively. We fit the spectral energy distribution and the MIR visibilities with radiative transfer simulations.}{We are able to reproduce the spectral energy distribution as well as the MIR visibility for one of the three epochs (third epoch) with a basic disk model. We were able to reproduce the very different visibility curve obtained nine years earlier with a very similar baseline (first epoch), using the same disk model with a smaller scale height. The same density distribution also reproduces the observation made with a higher spatial resolution in the second epoch, i.e. only two months before the third epoch.}{}


\bibpunct{(}{)}{;}{a}{}{,}

\begin{document}
    \maketitle
    
    \section{Introduction}
    \label{sec:intro}
        Young stellar objects (YSOs) such as T Tauri stars gather material from their surrounding disk via accretion. It is commonly believed that the observed brightness variability of YSOs are due to a temporal increase of the stellar accretion rate that only lasts a limited amount of time \citep{hartmann-kenyon-1996,mosoni-et-al-2013}. Various explanations for the temporal variations of accretion have been proposed: planet-disk interaction \citep{lodato-clarke-2004,ruge-et-al-2014}, thermal or gravitational instability \citep{pringle-1981,hartmann-kenyon-1996,lodato-rice-2004,lodato-rice-2005,zhu-et-al-2009}, interactions between the disk and a central binary \citep{pfalzner-2008}, and the impact of the magnetic field \citep{dangelo-spruit-2010}. The flux variations can be traced from ultraviolet (UV) up to mid-infrared (MIR) wavelengths has been the subject of many photometric and spectroscopic studies in the past \citep{lorenzetti-et-al-2009,semkov-et-al-2013,antoniucci-et-al-2014,banzatti-et-al-2014}. With the rise of optical/infrared long baseline interferometry, one is potentially able to put stringent constraints on the region of the origin of the flux variations, even if not to spatially resolve it.
        
        In this study, we investigate the variability of the protoplanetary disk of DR Tau, a classical T Tauri star. DR Tau is a member of the Taurus-Auriga molecular cloud at a distance of about 140~pc \citep{kenyon-et-al-1994}. It is known for its very strong, long- and short-term photometric and spectroscopic variability from the UV to the MIR \citep{kenyon-hartmann-1995,alencar-et-al-2001}. Therefore the obtained parameters of the surrounding disk and the star itself vary with every new study. For instance, \citet{banzatti-et-al-2014} showed that the accretion luminosity can change by a factor of about two to three within less than two months by analysing the UV excess. Besides, age and spectral type of DR Tau could only be limited to be less than 4~Myr and between K5 and K7, respectively \citep{muzerolle-et-al-2003,greaves-2004,mohanty-et-al-2005,pontoppidan-et-al-2011}.
        
        We present self-consistent disk models, simulated with the radiative transfer code MC3D \citep{wolf-et-al-1999,wolf-2003}. These models reproduce highly spatially resolved interferometric data obtained with the MID-infrared Interferometric instrument (MIDI) at the VLTI at three different epochs. The measured visibilities show a clear change in the brightness distribution of the disk and, thus, we are able to determine the underlying variation of the dust density distribution. To reduce the degeneracies, a set of published photometric data is also used to model the spectral energy distribution (SED).
        
    \section{Mid-infrared interferometric observations and data reduction}
    \label{sec:obs}
        
        The circumstellar disk of DR Tau was observed with MIDI \citep{leinert-et-al-2003a,leinert-et-al-2003b,morel-et-al-2004} at the VLTI at three epochs. For all measurements the instrument was fed by two 8.2~m Unit Telescopes and the prism was used to disperse the light with a low resolution of about 30.
        
        The data were reduced with the MIA+EWS package\footnote{\url{http://www.strw.leidenuniv.nl/~nevec/MIDI/}}. The visibilities shown were computed with MIA that is based on the analysis of the power spectrum. The visibility shown for the first night is the mean of the two calibrated measurements. The standard error is 0.07 on average. All visibilities have been calibrated by all the calibrators taken in the same night and with the same mode as the science target. The stability of the transfer function is reflected by the errors shown together with the visibilities. Calibrators with peculiar visibilities have been ignored and are indicated in the journal of observations; see \prettyref{tab:obs_data}. A comparison of the MIA results with the results obtained with EWS confirm the visibilities. For the last night the EWS visibility even tends to be lower up to about 0.1 at short N-band wavelengths. The EWS is based on a coherent analysis of the interferometric fringe signal \citep{jaffe-2004}. The differential phases of our observations do not significantly differ from $0\degr$. The maximum differential phase is at about $10\degr$ for E2 but strongly depends on the used calibrator and thus the uncertainties are very high. Besides, differential phases equal to zero do not necessarily mean that there are no asymmetries in the target because linear phase terms cannot be probed \citep{ratzka-et-al-2009}.
        
\begin{figure}
            \resizebox{\hsize}{!}{\includegraphics{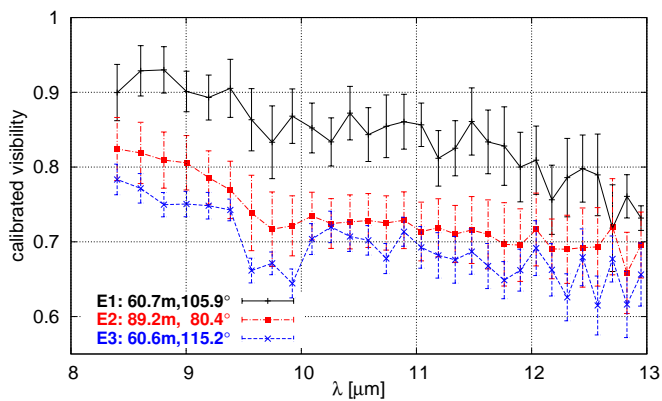}}
            \caption{Calibrated MIR visibility of DR Tau for the three different epochs January 2005 (E1), October 2013 (E2), and December 2013 (E3).}
            \label{fig:measured-vis}
        \end{figure}

        The data used for our analysis were taken in January 2005, October 2013, and December 2013. Additional measurements were made in October and November 2004 and in December 2013, but were excluded due to their low quality. The data set presented and modelled by \citet{schegerer-et-al-2009} has been re-reduced. The calibrated visibilities for the three epochs are shown in \prettyref{fig:measured-vis}.
        
        \begin{table*}
            \caption{Journal of our VLTI/MIDI observations. The length and position angle of the projected baselines have been determined from the fringe tracking sequence.}
            \label{tab:obs_data}
            \centering
            \begin{tabular}{c c c c c c c c}
                \hline\hline
                \rule{0pt}{2.5ex}Date of & Universal & Object & N-band flux & \multicolumn{2}{c}{proj. Baseline} & Interferometric & Photometric \\
                observation & time & & [Jy] & [m] & [deg] & frames & frames \\[1mm]
                \hline
                \multicolumn{8}{c}{\rule{0pt}{2.5ex}U3-U4 (high sensitivity / prism)} \\
                \multicolumn{8}{c}{\rule{0pt}{2.5ex}E1} \\[1mm]
                \hline
                \rule{0pt}{2.5ex}01-01-2005 & 01:42 - 01:54 & HD 37160\tablefootmark{a}\ \tablefootmark{b} & 6.5 & 60.4 & 112.5 & 12000$\times$16 ms & 2$\times$4000$\times$16 ms \\
                01-01-2005 & 02:41 - 02:48 & DR Tau   & -   & 61.0 & 106.1 & 8000$\times$12~ms  & 2$\times$1500$\times$12~ms \\
                01-01-2005 & 02:51 - 03:00 & DR Tau   & -   & 60.4 & 105.6 & 8000$\times$12~ms  & 2$\times$4000$\times$12~ms \\
                01-01-2005 & 03:19 - 03:26 & HD 31421 & 9.4 & 59.3 & 105.7 & 8000$\times$12~ms  & 2$\times$1500$\times$12~ms \\
                01-01-2005 & 03:59 - 04:07 & HD 31421 & 9.4 & 55.2 & 104.4 & 8000$\times$12~ms  & 2$\times$1500$\times$12~ms \\
                01-01-2005 & 05:24 - 05:31 & HD 49161 & 7.2 & 58.7 & 107.2 & 8000$\times$12~ms  & 2$\times$1500$\times$12~ms \\[1mm]
                \hline
                \multicolumn{8}{c}{\rule{0pt}{2.5ex}U2-U4 (high sensitivity / prism)} \\
                \multicolumn{8}{c}{\rule{0pt}{2.5ex}E2} \\[1mm]
                \hline
                \rule{0pt}{2.5ex}20-10-2013 & 07:10 - 07:54 & HD 25604 & 5.1 & 89.4 & 82.3 & 8000$\times$18~ms & 2$\times$4000$\times$18~ms \\
                20-10-2013 & 07:45 - 07:52 & HD 37160\tablefootmark{a} & 6.5 & 86.7 & 84.2 & 8000$\times$18~ms & 2$\times$4000$\times$18~ms \\
                20-10-2013 & 08:01 - 08:08 & HD 31421 & 9.4 & 89.4 & 82.1 & 8000$\times$18~ms & 2$\times$2000$\times$18~ms \\
                20-10-2013 & 08:16 - 08:24 & DR Tau   & -   & 89.2 & 80.4 & 8000$\times$18~ms & 2$\times$4000$\times$18~ms \\
                20-10-2013 & 08:33 - 08:41 & HD 69142 & 5.7 & 82.0 & 55.5 & 8000$\times$18~ms & 2$\times$4000$\times$18~ms \\[1mm]
                \hline
                \multicolumn{8}{c}{\rule{0pt}{2.5ex}U3-U4 (high sensitivity / prism)} \\
                \multicolumn{8}{c}{\rule{0pt}{2.5ex}E3} \\[1mm]
                \hline
                \rule{0pt}{2.5ex}20-12-2013 & 01:02 - 01:08 & HD 31421 & 9.4 & 57.2 & 117.9 & 8000$\times$18~ms & 2$\times$2000$\times$18~ms \\
                20-12-2013 & 01:30 - 01:37 & DR Tau   & -                    & 60.6 & 115.2 & 8000$\times$18~ms & 2$\times$4000$\times$18~ms \\
                20-12-2013 & 01:54 - 02:03 & HD 33554 & 7.7\tablefootmark{c} & 60.4 & 115.0 & 8000$\times$18~ms & 2$\times$6000$\times$18~ms \\
                20-12-2013 & 02:55 - 03:03 & HD 25604\tablefootmark{b} & 5.1 & 60.1 & 104.3 & 8000$\times$18~ms & 2$\times$4000$\times$18~ms \\
                20-12-2013 & 03:41 - 03:48 & HD 37160\tablefootmark{a} & 6.5 & 62.4 & 108.9 & 8000$\times$18~ms & 2$\times$4000$\times$18~ms \\
                \hline
            \end{tabular}
            \tablefoot{\tablefoottext{a}{Spectrophotometric calibrator;}
                \tablefoottext{b}{peculiar instrumental visibility;}
                \tablefoottext{c}{N-band flux taken from CalVin}}
        \end{table*}
        
    \section{Data analysis and disk modelling}
        The properties of our target DR Tau are listed in \prettyref{tab:prop_data}. Previous photometric measurements are given in \prettyref{tab:sed_data}.
            
        \begin{table}
            \caption{Stellar properties of DR Tau.}
            \label{tab:prop_data}
            \centering
            \begin{tabular}{l l r}
                \hline\hline
                \rule{0pt}{2.5ex}Parameter & Value                        & Ref. \\[1mm]
                \hline
                \rule{0pt}{2.5ex}RA (J2000)& 04h~47m~06\fs2               & 1 \\
                Dec (J2000)                & +16\degr~58\arcmin~42\farcs8 & 1 \\
                Distance                   & 140~pc                       & 2 \\
                Visual extinction          & 1.6~mag                      & 3 \\
                Spectral type              & K5--7                        & 3, 4 \\
                Stellar mass               & 0.4--1~$\text{M}_{\sun}$     & 4, 5 \\
                Total luminosity           & 1--5.5~$\text{L}_{\sun}$     & 3, 5, 6, 7 \\
                Age                        & $\lesssim$4~Myr              & 8, 9 \\
                \hline
            \end{tabular}
            \tablebib{
            (1)~2MASS All Sky Catalog of point sources \citep{cutri-et-al-2003}; (2) \citet{kenyon-et-al-1994}; (3) \citet{muzerolle-et-al-2003}; (4) \citet{pontoppidan-et-al-2011}; (5) \citet{isella-carpenter-sargent-2009}; (6) \citet{banzatti-et-al-2014}; (7) \citet{eisner-et-al-2014}; (8) \citet{mohanty-et-al-2005}; (9) \citet{greaves-2004}}
        \end{table}
        
        \begin{table}
            \caption{Photometric data of DR Tau.}
            \label{tab:sed_data}
            \centering
            \begin{tabular}{l c r}
                \hline\hline
                \rule{0pt}{2.5ex}$\lambda$ [\textmu m] & Flux [Jy] & Ref. \\[1mm]
                \hline
                \rule{0pt}{2.5ex}0.36 & 0.050 $\pm$ 0.033 & 1 \\
                0.44 & 0.061 $\pm$ 0.035 & 1 \\
                0.55 & 0.095 $\pm$ 0.047 & 1 \\
                0.64 & 0.149 $\pm$ 0.076 & 1 \\
                0.78 & 0.26  $\pm$ 0.15  & 1 \\
                1.25 & 0.46  $\pm$ 0.01  & 2 \\
                1.65 & 0.78  $\pm$ 0.04  & 2 \\
                2.17 & 1.22  $\pm$ 0.02  & 2 \\
                3.6  & 1.86  $\pm$ 0.20  & 3 \\
                4.5  & 1.89  $\pm$ 0.15  & 3 \\
                5.8  & 2.34  $\pm$ 0.02  & 4 \\
                8    & 1.92  $\pm$ 0.01  & 4 \\
                12   & 3.16  $\pm$ 0.03  & 5 \\
                25   & 4.30  $\pm$ 0.05  & 5 \\
                60   & 5.51  $\pm$ 0.04  & 5 \\
                100  & 5.73  $\pm$ 0.63  & 6 \\
                200  & 4.10  $\pm$ 0.85  & 7 \\
                450  & 2.38  $\pm$ 0.17  & 6 \\
                600  & 0.61  $\pm$ 0.05  & 8 \\
                729  & 0.40  $\pm$ 0.08  & 9 \\
                850  & 0.533 $\pm$ 0.007 & 6 \\
                1056 & 0.23  $\pm$ 0.02  & 9 \\
                1300 & 0.109 $\pm$ 0.011 & 10\\
                \hline
            \end{tabular}
            \tablebib{
            (1)~\citet{kenyon-hartmann-1995}; (2) 2MASS All Sky Catalog of point sources \citep{cutri-et-al-2003}; (3) \citet{robitaille-et-al-2007}; (4) \citet{hartmann-et-al-2005}; (5) \citet{weaver-jones-1992}; (6) \citet{andrews-williams-2005}; (7) ISO Data Archive; (8) \citet{mannings-emerson-1994}; (9) \citet{beckwith-sargent-1991}; (10) \citet{isella-carpenter-sargent-2009}}
        \end{table}
        
        \subsection{Temporal variability}
        \label{sec:temporal}
            
            \textit{Interferometry}: The visibilities obtained at the first epoch in January 2005 (E1\footnote{In the following, we denote the epochs 1, 2, and 3 as E1, E2, and E3, respectively.}) differ significantly from those made at E3 in December 2013, although the projected baseline lengths (BL) and orientations are nearly the same. These observations indicate changes of the brightness distribution and, hence, the underlying density distribution in the innermost regions of the disk ($\la$4~AU). Thus, these two observations cannot be reproduced with a static disk model. Possible reasons could be a general brightness increase of the disk due to a variable accretion rate onto the central star or azimuthal brightness asymmetries that have significantly changed their position between both epochs of observation. Local brightness asymmetries on the scale of several dozens to hundreds of AU have already been observed in the case of various protoplanetary disks, e.g. GM~Aur \citep{schneider-et-al-2003}, FS~Tau \citep{hioki-et-al-2011}, IM Lupi \citep{pinte-et-al-2008}. However, in contrast to these observations only long-baseline interferometry in the optical/infrared provides the angular resolution so far to trace regions, so close to the central star that the orbital motion on the scale of months to years has to be taken into account in the data analysis. In the very near future, ALMA will provide a similar angular resolution in the submm/mm wavelength range.\\
            On the other hand, the visibility obtained at E2 (October 2013) with the much longer BL of about 90~m is higher than the measurement with BL $\approx 60$~m two months later (E3). A higher visibility means more compact structure in the case of simple structures. Thus, an observation with a longer BL, i.e. tracing higher spatial frequencies, result in a lower visibility, which is not the case here. Of course, we observed at two different baseline orientations, which at first glance might explain this counterintuitive scenario. A static, rotationally symmetric, inclined disk, observed at slightly different orientations can produce different visibilities if one observation is almost aligned with the minor or major axis of the disk. This difference becomes larger with higher orientation difference and higher disk inclination, however, our observations were made at an orientation difference of only $35\degr$ and there is no evidence of a disk inclination larger than $60\degr$ \citep{kenyon-et-al-1994,alencar-et-al-2001,brown-et-al-2013,eisner-et-al-2014}; see also \prettyref{sec:res}. Thus, it could either be a hint of short-term variations at the same timescale as the UV variations observed by \citet{banzatti-et-al-2014} or for a non-rotationally symmetric disk, i.e. disk structures seen at different phases of their orbit. In summary, our interferometric observations indicate the presence of a non-axisymmetric structure of the disk and/or temporal variations of the disk structure in mid-infrared bright regions of the disk.\\
            \textit{Photometry}: For a moderately inclined disk, we expect to observe a silicate emission feature at $\approx$10~\textmu m and a smooth SED without depressions or peaks. In contrast to the adjacent photometric data points, the Spitzer/IRAC data suggest a lack of flux at 8~\textmu m; see \prettyref{tab:sed_data}. Moreover, this is the same wavelength region MIDI is sensitive to, and the non-stable visibility suggests a non-stable flux. Thus, it is possible that this photometric Spitzer observation was made during a more quiet phase of DR Tau. Because of this, we exclude the IRAC data point from our fitting procedure.
        
        \subsection{Model set-up}
        \label{sec:mod}
            
            \textit{Earlier studies}: Many different values have been published for the orientation of the disk, described by the disk inclination $i$ and the disk position angle $PA$. The inclination ranges from $9\degr$ \citep{pontoppidan-et-al-2011} to $60\degr$ \citep{akeson-et-al-2005} and the position angle is presumed to be between $75\degr$ \citep{eisner-et-al-2014} and $180\degr$ \citep{andrews-williams-2005}. The disk is rather massive, $M_{\text{disk}}\approx0.01$--$0.1~M_{\sun}$ \citep{akeson-et-al-2005,robitaille-et-al-2007,schegerer-et-al-2009,ricci-et-al-2010}, and the inner radius has been determined to be less than $0.1$~AU through analysing the emission of CO and H$_2$O \citep{brown-et-al-2013,banzatti-et-al-2014,eisner-et-al-2014}.\\
            \textit{Disk structure}: \citet{schegerer-et-al-2009} also fit photometric and interferometric data for DR Tau but only for one epoch (E1). Their results are used as a starting point for our simulation, which also applies to the Ansatz of \citet{shakura-sunyaev-1973} for the density distribution. It is given by
            \begin{equation}
                \varrho(r,z) = \varrho_0 \left(\frac{R_{\star}}{r}\right)^{\alpha}\exp{\left[-\frac{1}{2}\left(\frac{z}{h(r)}\right)^2\right]}\ ,
            \label{eq:density_dist}
            \end{equation}
            where $r$ and $z$ are the usual cylindrical coordinates, $R_{\star}$ the stellar radius, and $h(r)$ the scale height:
            \begin{equation}
                h(r) = h_{100}\left(\frac{r}{100\ \text{AU}}\right)^{\beta}\ .
            \label{eq:scale_height}
            \end{equation}
            The two exponents $\alpha$ and $\beta$ describe the radial density profile and disk flaring, respectively. Assuming a coupling of temperature and surface density, the number of free parameters decreases by one \citep{shakura-sunyaev-1973}:
            \begin{equation}
                \alpha = 3\left(\beta-\frac{1}{2}\right)\ .
            \label{eq:alpha_beta}
            \end{equation}
            This simple disk model has already been used successfully for fitting spatially resolved and unresolved multi-wavelength observations of several protoplanetary disks \citep{wolf-et-al-2003,ratzka-et-al-2009,sauter-wolf-2011,madlener-et-al-2012,grafe-et-al-2013,liu-et-al-2013}.\\
            \textit{Dust properties}: The dust in our set-up is a mixture of 62.5\% astronomical silicate and 37.5\% graphite (12.5\% parallel and 25\% perpendicular) and has particle sizes between 5~nm and 250~nm. The grain size distribution is given by the MRN distribution \citep{mathis-et-al-1977}, i.e. follows a power law, $n(s) \propto s^{-3.5}$, where $n\cdot$d$s$ are the number of particles in the radius interval $[s,s+\text{d}s]$.\\
            \textit{Heating sources}: The central star is treated as a black body with a fixed value for the total luminosity of $L_{\text{tot}} = L_{\star} + L_{\text{acc}} = 1.9~\text{L}_{\sun}$ \citep{muzerolle-et-al-2003,schegerer-et-al-2009,eisner-et-al-2014}. Further, we set the stellar effective temperature to $T_{\star} = 4050$~K \citep{mohanty-et-al-2005}. The disk around DR Tau is actively accreting \citep{robitaille-et-al-2007,edwards-et-al-2013,eisner-et-al-2014}. Therefore, we also assume an additional accretion heating of the disk. To implement this, we use a very simple, static accretion model, by adding a second point-like, black-body source in the centre of our disk. The properties of this accretion source, i.e. temperature and truncation radius, are fixed in this study to $T_{\text{accr}} = 8000$~K and $R_{\text{trunc}} = 5~R_{\star}$, respectively. These values were already successfully used to model other TTSs \citep{akeson-et-al-2005,schegerer-et-al-2008,schegerer-et-al-2009}.
            
            To calculate the dust temperature distribution and the corresponding SED and mid-infrared visibilities, we use the radiative-transfer code MC3D \citep{wolf-et-al-1999,wolf-2003}.
            
        \subsection{Modelling}
            
            \textit{Fitting procedure}: In the standard fitting approach, the photometric and interferometric data were used simultaneously. The SED is a degenerate quantity and more than just one adequate density distribution could reproduce the observations. The visibility curves are then used to reduce these degeneracies, since they are only sensitive to small-scale structures in the interior of the disk. According to \prettyref{sec:temporal}, this approach is not suitable for our observations. The visibilities obtained at the three epochs cannot be reproduced by a disk with only one parameter set. Thus, we independently fit all four sets, SED + three visibilities. Due to the set-up of our global model, a resulting parameter set based on a visibility is only valid for the innermost brightness distribution and does not represent the entire disk.\\
            \textit{Parameter space and fitness quality}: To explore the parameter space, we use an iteratively refined grid search. The boundaries of the parameter space and the overall number of used values for each parameter are shown in \prettyref{tab:param_space}. To compare all individual models, we calculate $\chi^2$ for each resulting SED
            
            \begin{equation}
                \chi^2 = \sum^{m}_{i=1}{\left(\frac{F_{\text{SED}}(\lambda_i)-F_{\text{model}}(\lambda_i)}{\sigma_{\text{SED}}(\lambda_i)}\right)^2}\ ,
            \end{equation}
            where $m$ is the number of wavelengths in \prettyref{tab:sed_data} and $\sigma_{\text{SED}}$ the photometric uncertainties. To evaluate the visibilities, we apply a similar approach. Because of the huge amount of time needed to calculate a radiation map with sufficient resolution, we do not calculate the visibilities and thus the $\chi^2$ value for every measured wavelength bin given by the spectral resolution of MIDI, but at seven wavelengths between 8 and 13~\textmu m.
            
            \begin{table}
                \caption{Boundaries of parameter space.}
                \label{tab:param_space}
                \centering
                \begin{tabular}{l l l r}
                    \hline\hline
                    \rule{0pt}{2.5ex}Disk parameter                  & Min. value       & Max. value       & n \\[1mm]
                    \hline
                    \rule{0pt}{2.5ex}$L_{\star}$~[$\text{L}_{\sun}$] & 0.25             & 1.9              & 9  \\
                    $L_{\text{accr}}$~[$\text{L}_{\sun}$]            & 0                & 1.65             & 9  \\
                    $M_{\text{dust}}$~[$\text{M}_{\sun}$]            & $8\times10^{-4}$ & $3\times10^{-3}$ & 2  \\
                    $R_{\text{in}}$~[AU]                             & 0.065            & 2                & 23 \\
                    $R_{\text{out}}$~[AU]                            & 70               & 350              & 5  \\
                    $\beta$                                          & 0.65             & 1.3              & 14 \\
                    $h_{\text{100}}$~[AU]                            & 10               & 20               & 13 \\
                    Inclination $i$~[$\degr$]                        & 8                & 62               & 14 \\
                    Position Angle $PA$~[$\degr$]                    & 0                & 175              & 36 \\
                    \hline
                \end{tabular}
            \end{table}
        
    \section{Results}
    \label{sec:res}
        
\begin{figure}
            \centering
            \resizebox{\hsize}{!}{\includegraphics{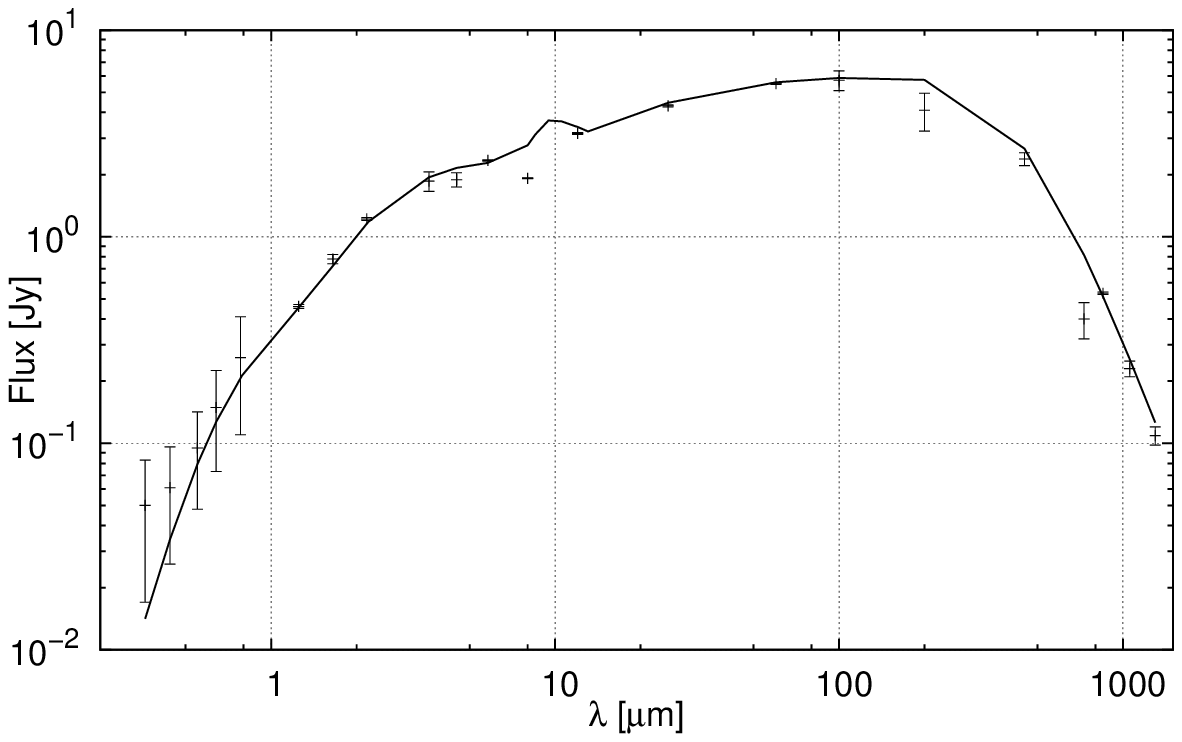}}\\
            \resizebox{\hsize}{!}{\includegraphics{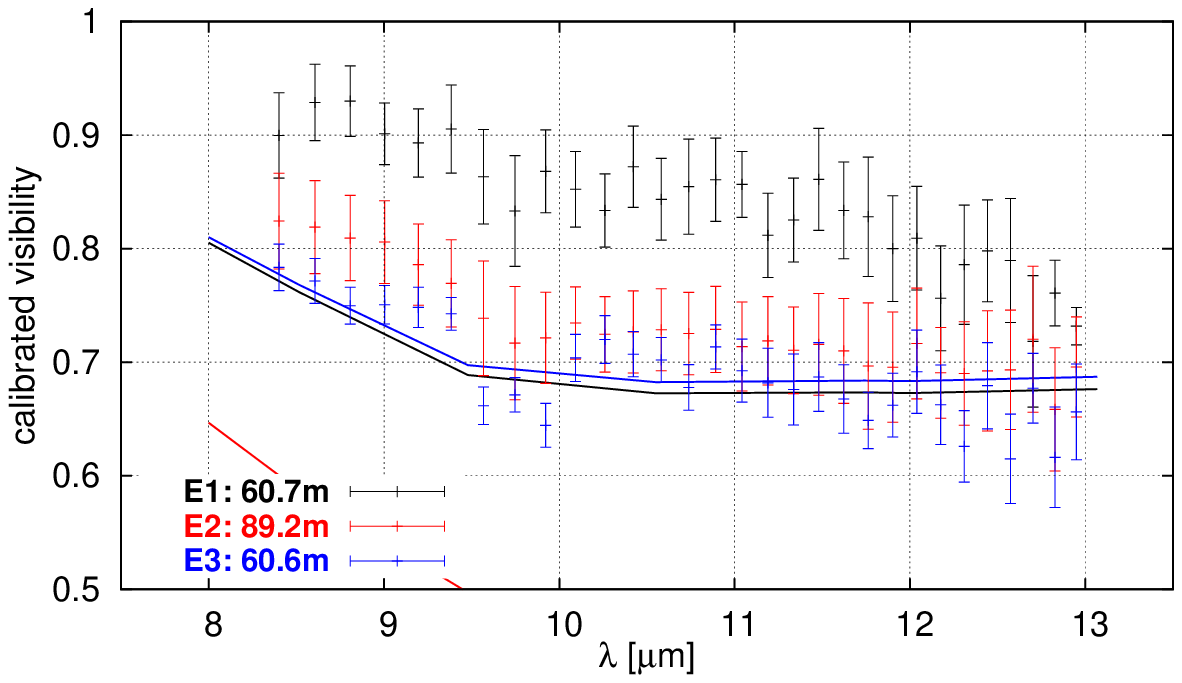}}
            \caption{Best SED fit with corresponding MIR visibilities.
            The three visibility epochs are shown in different colours (black: first, red: second, blue: third).
            Observations: with uncertainties; model: solid lines.}
        \label{fig:best_sed}
        \end{figure}

        In the upper part of \prettyref{fig:best_sed}, our best SED fit is shown. We excluded the Spitzer/IRAC data point at 8~\textmu m from the fitting process (see \prettyref{sec:temporal}). The properties of this disk are listed in the second column of \prettyref{tab:model-param}. These values are in agreement with other DR Tau studies; cf. \citet{banzatti-et-al-2014,eisner-et-al-2014,isella-carpenter-sargent-2009,schegerer-et-al-2009,muzerolle-et-al-2003}. These parameters, however, cover a rather wide range, which could be directly related to the short- and long-term variations of the disk structure. Interestingly, the best-fit parameters result in mid-infrared visibilities, which are in agreement with those obtained in E3, although the visibilities were not considered during the SED fitting process (see lower part of \prettyref{fig:best_sed}).
        
        \begin{table*}
            \caption{Model parameters for our best fits. In the last column, the results of \citet{schegerer-et-al-2009} are also printed for comparison.}
            \label{tab:model-param}
            \centering
            \begin{tabular}{l c c c c}
                \hline\hline
                \rule{0pt}{2.5ex}Disk parameter                  & \multicolumn{3}{c}{Best Fit Value} & \citet{schegerer-et-al-2009}\\[1mm]
                \hline
                \rule{0pt}{2.5ex}                                & SED   & E3                     & E1 + E2 & Model III\\
                \hline
                \rule{0pt}{2.5ex}$L_{\star}$~[$\text{L}_{\sun}$] & 0.9   & 0.9                    & 0.9 & 0.9 \\
                $L_{\text{accr}}$~[$\text{L}_{\sun}$]            & 1.0   & 1.0                    & 1.0 & 1.0 \\
                $T_{\star}$~[K]                                  & 4050 \tablefootmark{a} & 4050 \tablefootmark{a} & 4050 \tablefootmark{a} & 4050 \\
                $T_{\text{accr}}$~[K]                            & 8000 \tablefootmark{a} & 8000 \tablefootmark{a} & 8000 \tablefootmark{a} & 8000 \\
                $M_{\text{dust}}$~[$\text{M}_{\sun}$]            & 3$\times$10$^{-3}$ \tablefootmark{b} & 3$\times$10$^{-3}$ \tablefootmark{b} & 3$\times$10$^{-3}$ \tablefootmark{b} & 1$\times$10$^{-3}$ \\
                $R_{\text{in}}$~[AU]                             & 0.111 & 0.065\tablefootmark{c} & 0.065\tablefootmark{c} & 0.05 \\
                $R_{\text{out}}$~[AU]                            & 200   & 90                     & 350\tablefootmark{d} & 90 \\
                $\beta$                                          & 1.025 & 1.025                  & 1.025 & 0.75 \\
                $h_{\text{100}}$~[AU]                            & 18    & 17.5                   & 10\tablefootmark{c} & 15 \\
                Inclination $i$~[$\degr$]                        & 44    & 44                     & 44 & 20 \\
%                 Position Angle $PA$~[$\degr$]                    & 50    & 75                     & 15 \\
               \hline
            \end{tabular}
            \tablefoot{\tablefoottext{a}{fixed values;}
                       \tablefoottext{b}{only two values included;}
                       \tablefoottext{c}{lower boundary;}
                       \tablefoottext{d}{upper boundary}}
        \end{table*}
        
        The corresponding visibilities in \prettyref{fig:best_sed} are plotted for a $PA \approx 50\degr$ %PA on the sky is 90-Offset for major axis%
        from north to east for the major axis of the disk. As predicted in \prettyref{sec:temporal}, the two visibilities with nearly the same set-up from January 2005 (E1) and December 2013 (E3) cannot be reproduced with the same disk parameter values. Although the $PA$ is a free parameter here, the baseline orientation difference of just $10\degr$ is too small to produce this difference in the visibility. Because of the small inclination of the disk, the visibility is not very sensitive to changes in the $PA$. The calculated visibilities with $BL = 60.6$~m (blue solid line) vary only between 5\% and 12\% due to changes in the $PA$. This uncertainty is also confirmed by previous DR Tau studies: \citet{pontoppidan-et-al-2011} found a $PA$ of about $0\degr$, \citet{isella-carpenter-sargent-2009} found $100\degr$, \citet{brown-et-al-2013} published $140\degr$ and \citet{akeson-et-al-2005} derived $PA = 160\degr \pm 55\degr$.
        
\begin{figure}
            \centering
            \resizebox{\hsize}{!}{\includegraphics{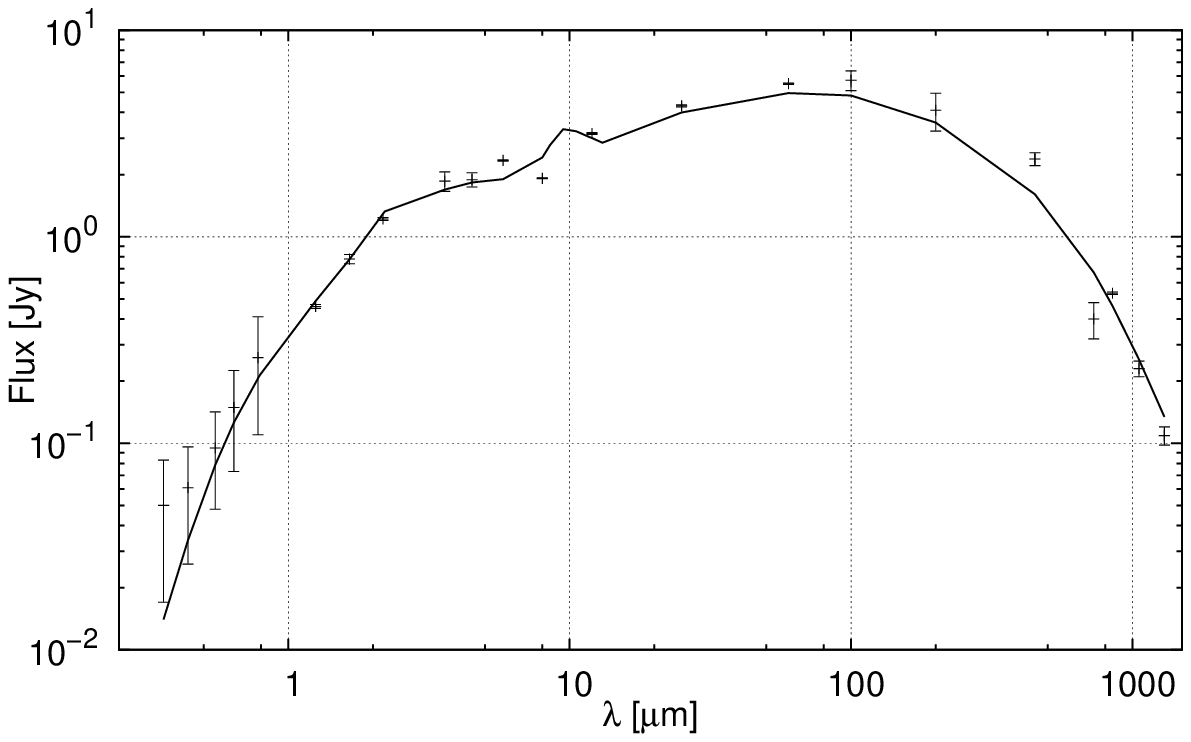}}\\
            \resizebox{\hsize}{!}{\includegraphics{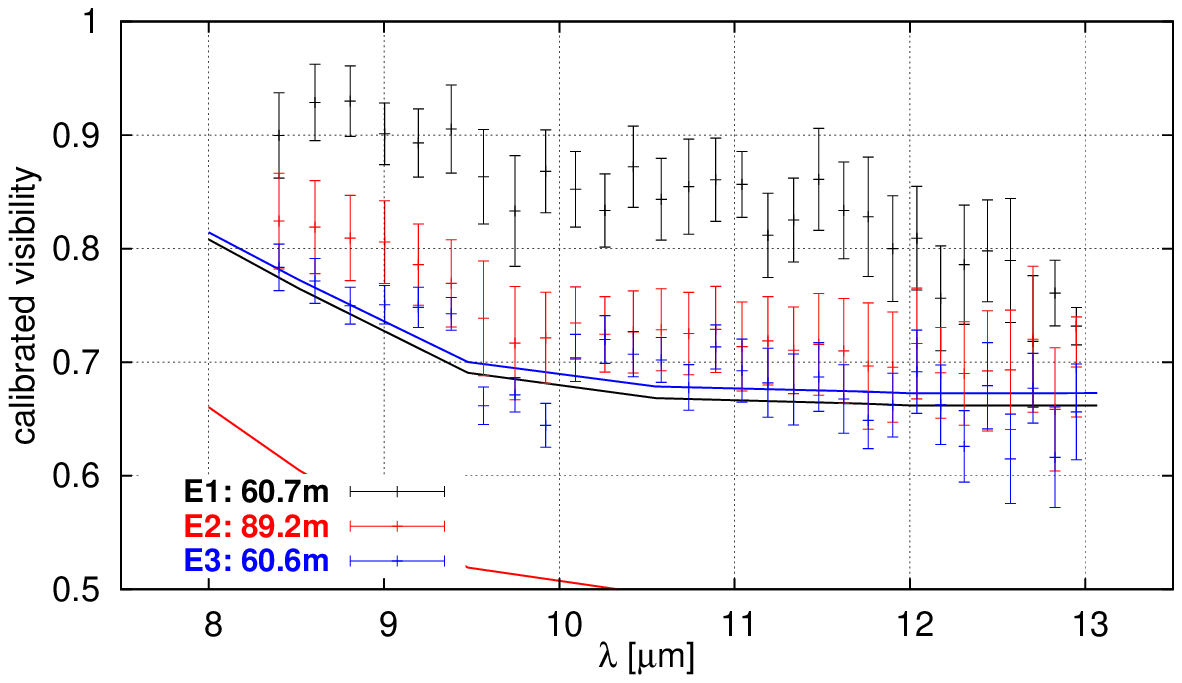}}
            \caption{Best fit for third epoch visibility with corresponding SED.
            The three visibility epochs are shown in different colours (black: first, red: second, blue: third).
            Observations: with uncertainties; model: solid lines.}
        \label{fig:best_vis_3}
        \end{figure}

        In \prettyref{fig:best_vis_3} our best result for the visibility obtained at epoch E3 is shown. As you can see, both the visibility curve for E3 and the SED are very similar to those in \prettyref{fig:best_sed}. The major differences in both set-ups are only in the inner radius, which changed from 0.111~AU to 0.065~AU, and the outer radius, which changed from 200~AU to 90~AU. The larger outer radius of the first model results in more cold dust, which increases the flux in the far-IR and (sub-)mm regime and matches the photometry better at these wavelengths. Furthermore, MIDI measurements are not sensitive to the outer disk because of the limited FoV of the interferometer. The other parameters are the same or very similar for both models, especially inclination and flaring, and thus the radial density profile is almost unchanged (see \prettyref{eq:alpha_beta}).
        
        A reasonable scenario for the found differences between the visibilities of E1 and E3 could be fast changes in the accretion luminosity, as mentioned in \prettyref{sec:intro}. To investigate this, we used the disk set-up that reproduces the SED best, and set $L_{\text{accr}} = 1.9~\text{L}_{\sun}$, resulting in a total luminosity of $L_{\text{tot}} = 2.8~\text{L}_{\sun}$. As a consequence, the dust heating is stronger at all distances and disk layers. This leads to a SED that is higher at all wavelengths and to visibilities that drop by about 10\% for all three baselines. If the accretion luminosity is decreased or even completely eliminated ($L_{\text{tot}} = L_{\star} = 0.9~\text{L}_{\sun}$), the visibilities increase up to 20\% for the longest wavelengths. Therefore, the lack of accretion cannot reproduce our observations as the visibilities are still too low for E1 and E2, and hence a different disk set-up is needed.
        
\begin{figure}
            \centering
            \resizebox{\hsize}{!}{\includegraphics{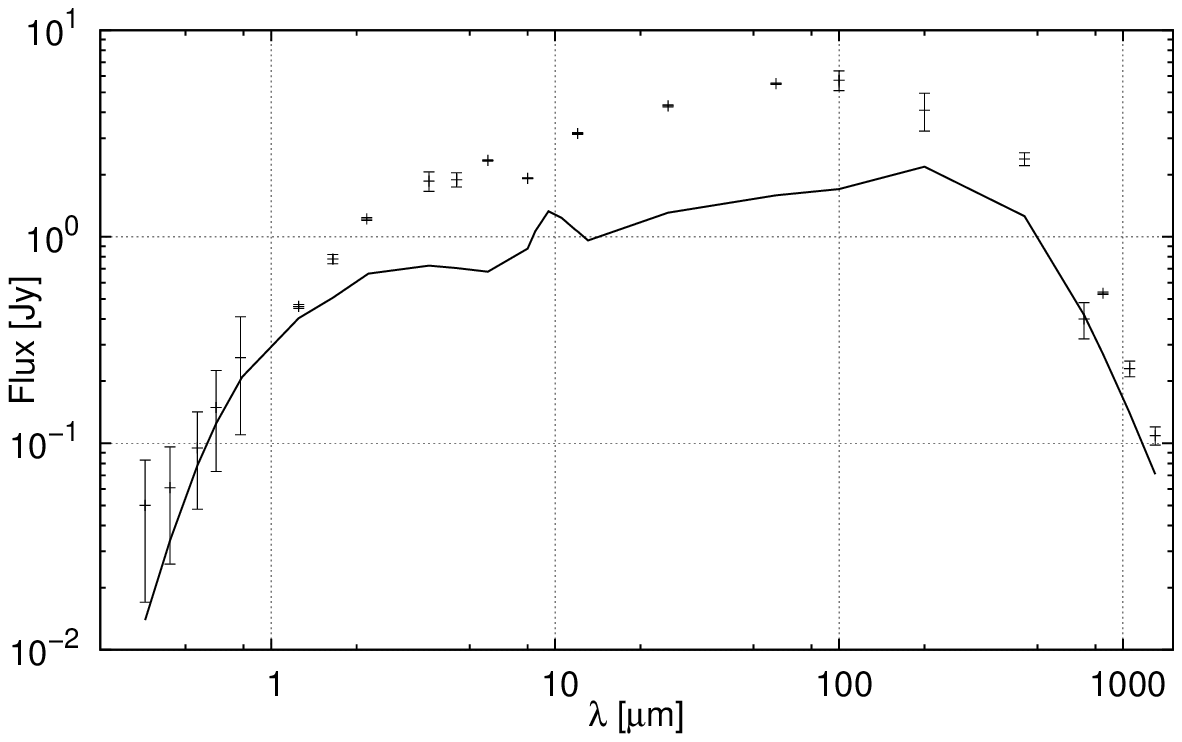}}\\
            \resizebox{\hsize}{!}{\includegraphics{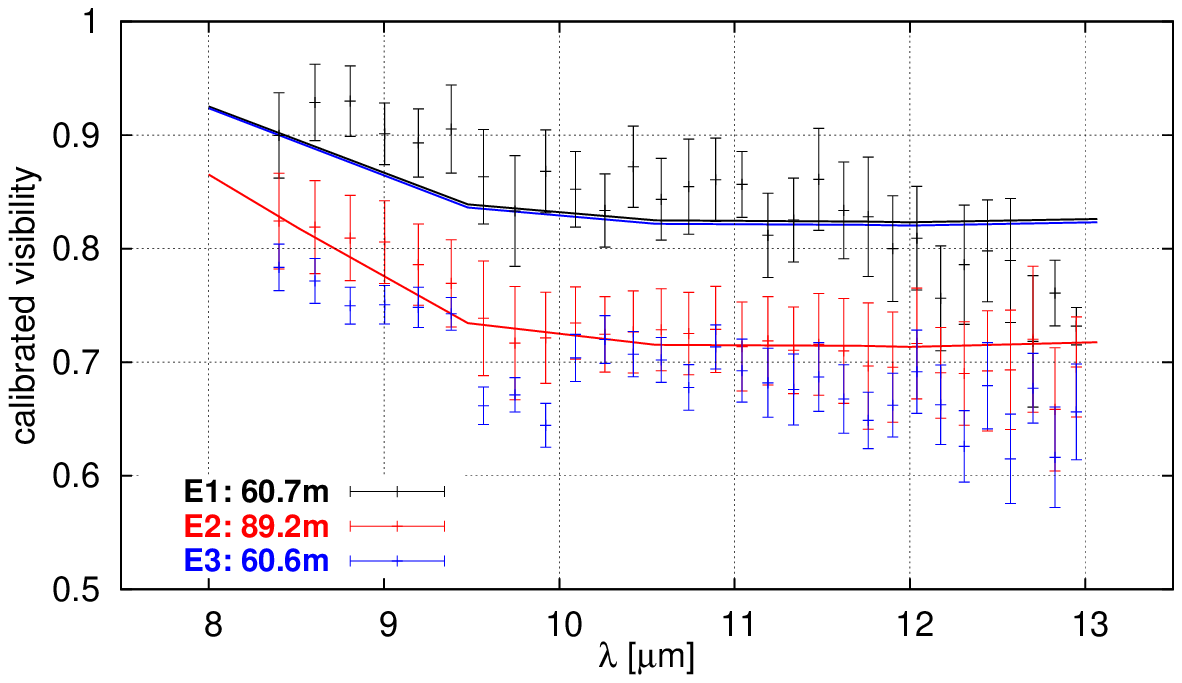}}
            \caption{Best fit for first and second epoch visibility with corresponding SED.
            The three visibility epochs are shown in different colours (black: first, red: second, blue: third).
            Observations: with uncertainties; model: solid lines.}
        \label{fig:best_vis_1_2}
        \end{figure}

        Our third parameter set reproduces the observations of E1 and E2 simultaneously, although they are almost nine years apart (see \prettyref{fig:best_vis_1_2}). If we compare the parameters of this best-fit model with those of the two previous models, we find that only the scale height and outer radius have changed and that flaring and inclination are again the same. This disk set-up is only valid for small stellar separations of $r \lesssim 10$~AU. More distant regions are not covered by the MIR visibilities and hence not considered by the fitting process. Because of the huge outer radius of model 3, the density in the hot, inner region is lower. In contrast, the scale height is decreased by nearly 50\%, which significantly increases the density in the midplane of the disk, resulting in a higher optical depth and a more compact emission. Hence, the visibility in the MIR is higher compared to the other two models and fits the observations. Owing to our purely global disk model, the SED in \prettyref{fig:best_vis_1_2} does not reproduce the observations and is too low for nearly all wavelengths. With the much smaller scale height, the disk of model 3 is very compact, resulting in a higher optical depth. Thus, the dust heating is less efficient, resulting in a reduced flux.
        
    \section{Conclusion}
    \label{sec:conclusion}
        
        We presented MIR visibilities of the T Tauri star DR Tau obtained with MIDI/VLTI. We modelled the SED and the individual mid-infrared visibility curves at three epochs to constrain the underlying variation in the disk density structure. Our simulations suggest that the hot, inner regions of the disk were more compact in January 2005 (E1) than they were in December 2013 (E3), especially in the layers close to the midplane. Since these two observations were made with nearly the same baseline length and position angle, it is possible that the structural difference is due to a local density variation, such as a density clump observed at different stages of its orbit. If this is true, the relevant region can be approximated by assuming a Keplerian orbit. If this hypothetical clump revolved once or twice within these nine years, its orbit must be about 2.5 or 4~AU, respectively. For a stellar distance of 140~pc, this is exactly the resolution of MIDI with a $BL \approx 60$~m. Since we cannot distinguish between $PA = 0\degr$ and $PA = 180\degr$, we detect the same fringe pattern twice per orbit. Although it is very unlikely, it cannot be completely ruled out that we observed the same density structure at first and second epoch. Considering a circular orbit of 2.5~AU with a period of $\approx4.5$~years, in two months the clump would have moved $\approx 13\degr-14\degr$ on its orbit. Together with the PA difference of $35\degr$ between E2 and E3, it is also possible to explain the non-detection at E3.
        
        However, we are aware of the fact that the photometric data set (see \prettyref{tab:sed_data}) is based on observations made at different epochs, i.e. at different stages of temporal evolution of the disk. Detailed studies over different timescales are necessary to achieve a better understanding of the temporal variations of protoplanetary disks and thus the underlying disk physics. With this pioneering study, we want to stress this problem, which has been commonly ignored in most disk modelling studies so far.
        
        With the second generation instrument MATISSE on the VLTI, a much better uv-coverage will be possible than with MIDI \citep{lopez-et-al-2014}. The MATISSE instrument will measure up to six visibilities at once with four telescopes, which is crucial for a consistent view on the disk at one certain point in time. Moreover, studying the temporal evolution of substructures could constrain physical processes in the potential planet-forming region \citep{kley-et-al-2009,flock-et-al-2015,bitsch-et-al-2015}. With existing MIDI data, new observations will enable us to trace disk evolution on timescales of more than 15~years. Even though MATISSE will cover a much wider wavelength range than MIDI (L, M, and N band), it is mandatory to observe protoplanetary disks with other interferometric instruments in different wavelength regimes also. High-spatial-resolution observations in the NIR and (sub-)mm, such as obtained with PIONIER/VLTI, AMBER/VLTI, and ALMA, will allow us to constrain the density distribution for dust at different temperatures, i.e. different disk regions.
        
    \begin{acknowledgement}
        
         This research was funded through the DFG grant: WO 857/13-1. This publication makes use of data products from the Two Micron All Sky Survey, which is a joint project of the University of Massachusetts and the Infrared Processing and Analysis Center/California Institute of Technology, funded by the National Aeronautics and Space Administration and the National Science Foundation.
        
    \end{acknowledgement}
    
    \bibliographystyle{aa}
    \bibliography{lit}

\begin{thebibliography}{57}
\expandafter\ifx\csname natexlab\endcsname\relax\def\natexlab#1{#1}\fi

\bibitem[{{Akeson} {et~al.}(2005){Akeson}, {Walker}, {Wood}, {Eisner}, {Scire},
  {Penprase}, {Ciardi}, {van Belle}, {Whitney}, \&
  {Bjorkman}}]{akeson-et-al-2005}
{Akeson}, R.~L., {Walker}, C.~H., {Wood}, K., {et~al.} 2005, \apj, 622, 440

\bibitem[{{Alencar} {et~al.}(2001){Alencar}, {Johns-Krull}, \&
  {Basri}}]{alencar-et-al-2001}
{Alencar}, S.~H.~P., {Johns-Krull}, C.~M., \& {Basri}, G. 2001, \aj, 122, 3335

\bibitem[{{Andrews} \& {Williams}(2005)}]{andrews-williams-2005}
{Andrews}, S.~M. \& {Williams}, J.~P. 2005, \apj, 631, 1134

\bibitem[{{Antoniucci} {et~al.}(2014){Antoniucci}, {Giannini}, {Li Causi}, \&
  {Lorenzetti}}]{antoniucci-et-al-2014}
{Antoniucci}, S., {Giannini}, T., {Li Causi}, G., \& {Lorenzetti}, D. 2014,
  \apj, 782, 51

\bibitem[{{Banzatti} {et~al.}(2014){Banzatti}, {Meyer}, {Manara},
  {Pontoppidan}, \& {Testi}}]{banzatti-et-al-2014}
{Banzatti}, A., {Meyer}, M.~R., {Manara}, C.~F., {Pontoppidan}, K.~M., \&
  {Testi}, L. 2014, \apj, 780, 26

\bibitem[{{Beckwith} \& {Sargent}(1991)}]{beckwith-sargent-1991}
{Beckwith}, S.~V.~W. \& {Sargent}, A.~I. 1991, \apj, 381, 250

\bibitem[{{Bitsch} {et~al.}(2015){Bitsch}, {Johansen}, {Lambrechts}, \&
  {Morbidelli}}]{bitsch-et-al-2015}
{Bitsch}, B., {Johansen}, A., {Lambrechts}, M., \& {Morbidelli}, A. 2015, \aap,
  575, A28

\bibitem[{{Brown} {et~al.}(2013){Brown}, {Troutman}, \&
  {Gibb}}]{brown-et-al-2013}
{Brown}, L.~R., {Troutman}, M.~R., \& {Gibb}, E.~L. 2013, \apjl, 770, L14

\bibitem[{{Cutri} {et~al.}(2003){Cutri}, {Skrutskie}, {van Dyk}, {Beichman},
  {Carpenter}, {Chester}, {Cambresy}, {Evans}, {Fowler}, {Gizis}, {Howard},
  {Huchra}, {Jarrett}, {Kopan}, {Kirkpatrick}, {Light}, {Marsh}, {McCallon},
  {Schneider}, {Stiening}, {Sykes}, {Weinberg}, {Wheaton}, {Wheelock}, \&
  {Zacarias}}]{cutri-et-al-2003}
{Cutri}, R.~M., {Skrutskie}, M.~F., {van Dyk}, S., {et~al.} 2003, {2MASS All
  Sky Catalog of point sources.}

\bibitem[{{D'Angelo} \& {Spruit}(2010)}]{dangelo-spruit-2010}
{D'Angelo}, C.~R. \& {Spruit}, H.~C. 2010, \mnras, 406, 1208

\bibitem[{{Edwards} {et~al.}(2013){Edwards}, {Kwan}, {Fischer}, {Hillenbrand},
  {Finn}, {Fedorenko}, \& {Feng}}]{edwards-et-al-2013}
{Edwards}, S., {Kwan}, J., {Fischer}, W., {et~al.} 2013, \apj, 778, 148

\bibitem[{{Eisner} {et~al.}(2014){Eisner}, {Hillenbrand}, \&
  {Stone}}]{eisner-et-al-2014}
{Eisner}, J.~A., {Hillenbrand}, L.~A., \& {Stone}, J.~M. 2014, \mnras, 443,
  1916

\bibitem[{{Flock} {et~al.}(2015){Flock}, {Ruge}, {Dzyurkevich}, {Henning},
  {Klahr}, \& {Wolf}}]{flock-et-al-2015}
{Flock}, M., {Ruge}, J.~P., {Dzyurkevich}, N., {et~al.} 2015, \aap, 574, A68

\bibitem[{{Gr{\"a}fe} {et~al.}(2013){Gr{\"a}fe}, {Wolf}, {Guilloteau},
  {Dutrey}, {Stapelfeldt}, {Pontoppidan}, \& {Sauter}}]{grafe-et-al-2013}
{Gr{\"a}fe}, C., {Wolf}, S., {Guilloteau}, S., {et~al.} 2013, \aap, 553, A69

\bibitem[{{Greaves}(2004)}]{greaves-2004}
{Greaves}, J.~S. 2004, \mnras, 351, L99

\bibitem[{{Hartmann} \& {Kenyon}(1996)}]{hartmann-kenyon-1996}
{Hartmann}, L. \& {Kenyon}, S.~J. 1996, \araa, 34, 207

\bibitem[{{Hartmann} {et~al.}(2005){Hartmann}, {Megeath}, {Allen}, {Luhman},
  {Calvet}, {D'Alessio}, {Franco-Hernandez}, \& {Fazio}}]{hartmann-et-al-2005}
{Hartmann}, L., {Megeath}, S.~T., {Allen}, L., {et~al.} 2005, \apj, 629, 881

\bibitem[{{Hioki} {et~al.}(2011){Hioki}, {Itoh}, {Oasa}, {Fukagawa}, \&
  {Hayashi}}]{hioki-et-al-2011}
{Hioki}, T., {Itoh}, Y., {Oasa}, Y., {Fukagawa}, M., \& {Hayashi}, M. 2011,
  \pasj, 63, 543

\bibitem[{{Isella} {et~al.}(2009){Isella}, {Carpenter}, \&
  {Sargent}}]{isella-carpenter-sargent-2009}
{Isella}, A., {Carpenter}, J.~M., \& {Sargent}, A.~I. 2009, \apj, 701, 260

\bibitem[{{Jaffe}(2004)}]{jaffe-2004}
{Jaffe}, W.~J. 2004, in Society of Photo-Optical Instrumentation Engineers
  (SPIE) Conference Series, Vol. 5491, New Frontiers in Stellar Interferometry,
  ed. W.~A. {Traub}, 715

\bibitem[{{Kenyon} {et~al.}(1994){Kenyon}, {Dobrzycka}, \&
  {Hartmann}}]{kenyon-et-al-1994}
{Kenyon}, S.~J., {Dobrzycka}, D., \& {Hartmann}, L. 1994, \aj, 108, 1872

\bibitem[{{Kenyon} \& {Hartmann}(1995)}]{kenyon-hartmann-1995}
{Kenyon}, S.~J. \& {Hartmann}, L. 1995, \apjs, 101, 117

\bibitem[{{Kley} {et~al.}(2009){Kley}, {Bitsch}, \& {Klahr}}]{kley-et-al-2009}
{Kley}, W., {Bitsch}, B., \& {Klahr}, H. 2009, \aap, 506, 971

\bibitem[{{Leinert} {et~al.}(2003{\natexlab{a}}){Leinert}, {Graser},
  {Przygodda}, {Waters}, {Perrin}, {Jaffe}, {Lopez}, {Bakker}, {B{\"o}hm},
  {Chesneau}, {Cotton}, {Damstra}, {de Jong}, {Glazenborg-Kluttig}, {Grimm},
  {Hanenburg}, {Laun}, {Lenzen}, {Ligori}, {Mathar}, {Meisner}, {Morel},
  {Morr}, {Neumann}, {Pel}, {Schuller}, {Rohloff}, {Stecklum}, {Storz}, {von
  der L{\"u}he}, \& {Wagner}}]{leinert-et-al-2003a}
{Leinert}, C., {Graser}, U., {Przygodda}, F., {et~al.} 2003{\natexlab{a}},
  \apss, 286, 73

\bibitem[{{Leinert} {et~al.}(2003{\natexlab{b}}){Leinert}, {Graser}, {Waters},
  {Perrin}, {Jaffe}, {Lopez}, {Przygodda}, {Chesneau}, {Schuller},
  {Glazenborg-Kluttig}, {Laun}, {Ligori}, {Meisner}, {Wagner}, {Bakker},
  {Cotton}, {de Jong}, {Mathar}, {Neumann}, \& {Storz}}]{leinert-et-al-2003b}
{Leinert}, C., {Graser}, U., {Waters}, L.~B.~F.~M., {et~al.}
  2003{\natexlab{b}}, in Society of Photo-Optical Instrumentation Engineers
  (SPIE) Conference Series, Vol. 4838, Interferometry for Optical Astronomy II,
  ed. W.~A. {Traub}, 893--904

\bibitem[{{Liu} {et~al.}(2013){Liu}, {Wang}, {Wolf}, \&
  {Madlener}}]{liu-et-al-2013}
{Liu}, Y., {Wang}, H.-C., {Wolf}, S., \& {Madlener}, D. 2013, Research in
  Astronomy and Astrophysics, 13, 841

\bibitem[{{Lodato} \& {Clarke}(2004)}]{lodato-clarke-2004}
{Lodato}, G. \& {Clarke}, C.~J. 2004, \mnras, 353, 841

\bibitem[{{Lodato} \& {Rice}(2004)}]{lodato-rice-2004}
{Lodato}, G. \& {Rice}, W.~K.~M. 2004, \mnras, 351, 630

\bibitem[{{Lodato} \& {Rice}(2005)}]{lodato-rice-2005}
{Lodato}, G. \& {Rice}, W.~K.~M. 2005, \mnras, 358, 1489

\bibitem[{{Lopez} {et~al.}(2014){Lopez}, {Lagarde}, {Jaffe}, {Petrov},
  {Sch{\"o}ller}, {Antonelli}, {Beckmann}, {Berio}, {Bettonvil}, {Glindemann},
  {Gonzalez}, {Graser}, {Hofmann}, {Millour}, {Robbe-Dubois}, {Venema}, {Wolf},
  {Henning}, {Lanz}, {Weigelt}, {Agocs}, {Bailet}, {Bresson}, {Bristow},
  {Dugu{\'e}}, {Heininger}, {Kroes}, {Laun}, {Lehmitz}, {Neumann}, {Augereau},
  {Avila}, {Behrend}, {van Belle}, {Berger}, {van Boekel}, {Bonhomme},
  {Bourget}, {Brast}, {Clausse}, {Connot}, {Conzelmann}, {Cruzal{\`e}bes},
  {Csepany}, {Danchi}, {Delbo}, {Delplancke}, {Dominik}, {van Duin}, {Elswijk},
  {Fantei}, {Finger}, {Gabasch}, {Gay}, {Girard}, {Girault}, {Gitton},
  {Glazenborg}, {Gont{\'e}}, {Guitton}, {Guniat}, {De Haan}, {Haguenauer},
  {Hanenburg}, {Hogerheijde}, {ter Horst}, {Hron}, {Hugues}, {Hummel},
  {Idserda}, {Ives}, {Jakob}, {Jasko}, {Jolley}, {Kiraly}, {K{\"o}hler},
  {Kragt}, {Kroener}, {Kuindersma}, {Labadie}, {Leinert}, {Le Poole}, {Lizon},
  {Lucuix}, {Marcotto}, {Martinache}, {Martinot-Lagarde}, {Mathar}, {Matter},
  {Mauclert}, {Mehrgan}, {Meilland}, {Meisenheimer}, {Meisner}, {Mellein},
  {Menardi}, {Menut}, {Merand}, {Morel}, {Mosoni}, {Navarro}, {Nussbaum},
  {Ottogalli}, {Palsa}, {Panduro}, {Pantin}, {Parra}, {Percheron}, {Duc},
  {Pott}, {Pozna}, {Przygodda}, {Rabbia}, {Richichi}, {Rigal}, {Roelfsema},
  {Rupprecht}, {Schertl}, {Schmidt}, {Schuhler}, {Schuil}, {Spang},
  {Stegmeier}, {Thiam}, {Tromp}, {Vakili}, {Vannier}, {Wagner}, \&
  {Woillez}}]{lopez-et-al-2014}
{Lopez}, B., {Lagarde}, S., {Jaffe}, W., {et~al.} 2014, The Messenger, 157, 5

\bibitem[{{Lorenzetti} {et~al.}(2009){Lorenzetti}, {Larionov}, {Giannini},
  {Arkharov}, {Antoniucci}, {Nisini}, \& {Di Paola}}]{lorenzetti-et-al-2009}
{Lorenzetti}, D., {Larionov}, V.~M., {Giannini}, T., {et~al.} 2009, \apj, 693,
  1056

\bibitem[{{Madlener} {et~al.}(2012){Madlener}, {Wolf}, {Dutrey}, \&
  {Guilloteau}}]{madlener-et-al-2012}
{Madlener}, D., {Wolf}, S., {Dutrey}, A., \& {Guilloteau}, S. 2012, \aap, 543,
  A81

\bibitem[{{Mannings} \& {Emerson}(1994)}]{mannings-emerson-1994}
{Mannings}, V. \& {Emerson}, J.~P. 1994, \mnras, 267, 361

\bibitem[{{Mathis} {et~al.}(1977){Mathis}, {Rumpl}, \&
  {Nordsieck}}]{mathis-et-al-1977}
{Mathis}, J.~S., {Rumpl}, W., \& {Nordsieck}, K.~H. 1977, \apj, 217, 425

\bibitem[{{Mohanty} {et~al.}(2005){Mohanty}, {Jayawardhana}, \&
  {Basri}}]{mohanty-et-al-2005}
{Mohanty}, S., {Jayawardhana}, R., \& {Basri}, G. 2005, \apj, 626, 498

\bibitem[{{Morel} {et~al.}(2004){Morel}, {Ballester}, {Bauvir}, {Biereichel},
  {Cuby}, {Galliano}, {Haddad}, {Housen}, {Hummel}, {Kaufer}, {Kervella},
  {Percheron}, {Puech}, {Rantakyro}, {Richichi}, {Sabet}, {Schoeller},
  {Spyromilio}, {Vannier}, {Wallander}, {Wittkowski}, {Leinert}, {Graser},
  {Neumann}, {Jaffe}, \& {de Jong}}]{morel-et-al-2004}
{Morel}, S., {Ballester}, P., {Bauvir}, B., {et~al.} 2004, in Society of
  Photo-Optical Instrumentation Engineers (SPIE) Conference Series, Vol. 5491,
  New Frontiers in Stellar Interferometry, ed. W.~A. {Traub}, 1666

\bibitem[{{Mosoni} {et~al.}(2013){Mosoni}, {Sipos}, {{\'A}brah{\'a}m},
  {Mo{\'o}r}, {K{\'o}sp{\'a}l}, {Henning}, {Juh{\'a}sz}, {Kun}, {Leinert},
  {Quanz}, {Ratzka}, {Schegerer}, {van Boekel}, \& {Wolf}}]{mosoni-et-al-2013}
{Mosoni}, L., {Sipos}, N., {{\'A}brah{\'a}m}, P., {et~al.} 2013, \aap, 552, A62

\bibitem[{{Muzerolle} {et~al.}(2003){Muzerolle}, {Calvet}, {Hartmann}, \&
  {D'Alessio}}]{muzerolle-et-al-2003}
{Muzerolle}, J., {Calvet}, N., {Hartmann}, L., \& {D'Alessio}, P. 2003, \apjl,
  597, L149

\bibitem[{{Pfalzner}(2008)}]{pfalzner-2008}
{Pfalzner}, S. 2008, \aap, 492, 735

\bibitem[{{Pinte} {et~al.}(2008){Pinte}, {Padgett}, {M{\'e}nard},
  {Stapelfeldt}, {Schneider}, {Olofsson}, {Pani{\'c}}, {Augereau},
  {Duch{\^e}ne}, {Krist}, {Pontoppidan}, {Perrin}, {Grady}, {Kessler-Silacci},
  {van Dishoeck}, {Lommen}, {Silverstone}, {Hines}, {Wolf}, {Blake}, {Henning},
  \& {Stecklum}}]{pinte-et-al-2008}
{Pinte}, C., {Padgett}, D.~L., {M{\'e}nard}, F., {et~al.} 2008, \aap, 489, 633

\bibitem[{{Pontoppidan} {et~al.}(2011){Pontoppidan}, {van Dishoeck}, {Blake},
  {Smith}, {Brown}, {Herczeg}, {Bast}, {Mandell}, {Smette}, {Thi}, {Young},
  {Morris}, {Dent}, \& {K{\"a}ufl}}]{pontoppidan-et-al-2011}
{Pontoppidan}, K.~M., {van Dishoeck}, E., {Blake}, G.~A., {et~al.} 2011, The
  Messenger, 143, 32

\bibitem[{{Pringle}(1981)}]{pringle-1981}
{Pringle}, J.~E. 1981, \araa, 19, 137

\bibitem[{{Ratzka} {et~al.}(2009){Ratzka}, {Schegerer}, {Leinert},
  {{\'A}brah{\'a}m}, {Henning}, {Herbst}, {K{\"o}hler}, {Wolf}, \&
  {Zinnecker}}]{ratzka-et-al-2009}
{Ratzka}, T., {Schegerer}, A.~A., {Leinert}, C., {et~al.} 2009, \aap, 502, 623

\bibitem[{{Ricci} {et~al.}(2010){Ricci}, {Testi}, {Natta}, {Neri}, {Cabrit}, \&
  {Herczeg}}]{ricci-et-al-2010}
{Ricci}, L., {Testi}, L., {Natta}, A., {et~al.} 2010, \aap, 512, A15

\bibitem[{{Robitaille} {et~al.}(2007){Robitaille}, {Whitney}, {Indebetouw}, \&
  {Wood}}]{robitaille-et-al-2007}
{Robitaille}, T.~P., {Whitney}, B.~A., {Indebetouw}, R., \& {Wood}, K. 2007,
  \apjs, 169, 328

\bibitem[{{Ruge} {et~al.}(2014){Ruge}, {Wolf}, {Uribe}, \&
  {Klahr}}]{ruge-et-al-2014}
{Ruge}, J.~P., {Wolf}, S., {Uribe}, A.~L., \& {Klahr}, H.~H. 2014, \aap, 572,
  L2

\bibitem[{{Sauter} \& {Wolf}(2011)}]{sauter-wolf-2011}
{Sauter}, J. \& {Wolf}, S. 2011, \aap, 527, A27

\bibitem[{{Schegerer} {et~al.}(2009){Schegerer}, {Wolf}, {Hummel}, {Quanz}, \&
  {Richichi}}]{schegerer-et-al-2009}
{Schegerer}, A.~A., {Wolf}, S., {Hummel}, C.~A., {Quanz}, S.~P., \& {Richichi},
  A. 2009, \aap, 502, 367

\bibitem[{{Schegerer} {et~al.}(2008){Schegerer}, {Wolf}, {Ratzka}, \&
  {Leinert}}]{schegerer-et-al-2008}
{Schegerer}, A.~A., {Wolf}, S., {Ratzka}, T., \& {Leinert}, C. 2008, \aap, 478,
  779

\bibitem[{{Schneider} {et~al.}(2003){Schneider}, {Wood}, {Silverstone},
  {Hines}, {Koerner}, {Whitney}, {Bjorkman}, \&
  {Lowrance}}]{schneider-et-al-2003}
{Schneider}, G., {Wood}, K., {Silverstone}, M.~D., {et~al.} 2003, \aj, 125,
  1467

\bibitem[{{Semkov} {et~al.}(2013){Semkov}, {Peneva}, {Munari}, {Dennefeld},
  {Mito}, {Dimitrov}, {Ibryamov}, \& {Stoyanov}}]{semkov-et-al-2013}
{Semkov}, E.~H., {Peneva}, S.~P., {Munari}, U., {et~al.} 2013, \aap, 556, A60

\bibitem[{{Shakura} \& {Sunyaev}(1973)}]{shakura-sunyaev-1973}
{Shakura}, N.~I. \& {Sunyaev}, R.~A. 1973, \aap, 24, 337

\bibitem[{{Weaver} \& {Jones}(1992)}]{weaver-jones-1992}
{Weaver}, W.~B. \& {Jones}, G. 1992, \apjs, 78, 239

\bibitem[{{Wolf}(2003)}]{wolf-2003}
{Wolf}, S. 2003, Computer Physics Communications, 150, 99

\bibitem[{{Wolf} {et~al.}(1999){Wolf}, {Henning}, \&
  {Stecklum}}]{wolf-et-al-1999}
{Wolf}, S., {Henning}, T., \& {Stecklum}, B. 1999, \aap, 349, 839

\bibitem[{{Wolf} {et~al.}(2003){Wolf}, {Padgett}, \&
  {Stapelfeldt}}]{wolf-et-al-2003}
{Wolf}, S., {Padgett}, D.~L., \& {Stapelfeldt}, K.~R. 2003, \apj, 588, 373

\bibitem[{{Zhu} {et~al.}(2009){Zhu}, {Hartmann}, {Gammie}, \&
  {McKinney}}]{zhu-et-al-2009}
{Zhu}, Z., {Hartmann}, L., {Gammie}, C., \& {McKinney}, J.~C. 2009, \apj, 701,
  620

\end{thebibliography}

\end{document}